\documentclass[12pt]{article}

\usepackage{amssymb,esvect,amsmath,graphicx,latexsym,amsthm,slashed,eso-pic,hyperref}
\usepackage[final]{pdfpages}

\usepackage{epsfig,amsmath,amssymb,verbatim,mathrsfs,hyperref}
\usepackage{xspace}
\usepackage{xcolor}
\usepackage{slashed}
\usepackage{dcolumn}
\usepackage{multirow}
\graphicspath{{./}{Figs/}}

\def\beq{\begin{eqnarray}}
\def\eeq{\end{eqnarray}}
\def\bea{\begin{eqnarray}}
\def\eea{\end{eqnarray}}

\newcommand{\gsim}{\lower.7ex\hbox{$\;\stackrel{\textstyle>}{\sim}\;$}}
\newcommand{\lsim}{\lower.7ex\hbox{$\;\stackrel{\textstyle<}{\sim}\;$}}

\newcommand{\beg}{{^8\textrm{Be}}}
\newcommand{\bes}{{^8\textrm{Be}}^*}
\newcommand{\bet}{{^8\textrm{Be}}^{*\prime}}
\newcommand{\lig}{{^7\textrm{Li}}}

\definecolor{green1}{RGB}{0,105,0}
\definecolor{cyan1}{RGB}{0,250,250}
\definecolor{magenta1}{RGB}{250,0,250}

\begin{document}
\begin{titlepage}
\begin{flushright}
{\large 
ACFI-T17-19\\
}
\end{flushright}
\noindent
\begin{center}
  \begin{LARGE}
    \begin{bf}
Dark Photons from Nuclear Transitions
     \end{bf}
  \end{LARGE}
\end{center}
\vspace{0.3cm}
\begin{center}
\begin{large}
Jonathan Kozaczuk
\end{large}
\vspace{1cm}\\
\begin{it}
Amherst Center for Fundamental Interactions, Department of Physics,\\ 
University of Massachusetts, Amherst, MA 01003, USA

\vspace{0.3cm}
email: \emph{\texttt{kozaczuk@umass.edu}}, 

\vspace{0.2cm}
\end{it}
\end{center}
\center{\today}

\begin{abstract}

Light new particles can be emitted in decays of excited nuclear states. Experiments analyzing such transitions and incorporating high-resolution detectors can be sensitive to new MeV-scale physics at a level competitive with upcoming collider and other fixed target experiments, provided sufficient luminosity. We demonstrate this in the case of the $^8\textrm{Be}$ system, showing that searches targeting the reported anomaly in $^8\textrm{Be}$ nuclear transitions can also be sensitive to currently unexplored regions of the canonical dark photon parameter space with 1 MeV $\lesssim m_{A^{\prime}} \lesssim 18$ MeV and $\varepsilon^2 \gtrsim10^{-7}$.  These experiments could be performed on a short timescale, at low cost, and directly probe both the hadronic and leptonic couplings of light hidden particles.

\end{abstract}

\end{titlepage}

\setcounter{page}{2}


\section{Introduction\label{sec:intro}}

Observations of rare nuclear transitions can be used to search for new hidden particles at the MeV scale~\cite{Donnelly:1978ty, Treiman:1978ge}. In this approach, a fixed target is bombarded with a hadron beam to produce excited states of a nucleus. The excited state then decays, and in doing so can emit new particles. This allows for resonantly-enhanced production cross-sections of light hidden degrees of freedom with low beam energies ($\sim 0.1 - 1$ MeV), high luminosities, and with precise predictions for the final state kinematics.  Historically, nuclear decay experiments targeted axions in the 1-10 MeV mass range~\cite{Savage:1986ty, Hallin:1986gh, Savage:1988rg}, and set some of the earliest limits on light Higgs bosons~\cite{Freedman:1984sd}. With the advent of more powerful accelerators, the non-detection of an MeV-scale axion, and the limited mass range accessible to nuclear transitions relative to other approaches, however, these searches became less competitive as probes of new physics. 

Recently, intriguing results~\cite{Krasznahorkay:2015iga, Krasznahorkay:2017qfd, Krasznahorkay:2017gwn} from a nuclear transition experiment at the MTA Atomki facility have sparked a renewed interest in the possibility of producing hidden MeV-scale particles in such experiments. The Atomki group observes a bump-like feature in the angular distribution of $e^+e^-$ pairs produced in decays of excited states of $^8$Be. Their findings have generated much interest in the particle physics community (beginning with Ref.~\cite{Feng:2016jff}), since they suggest a light new particle coupling to hadrons and leptons. Clearly, this result warrants future study in a similar, dedicated experiment, and there are currently proposals for follow-up searches that could be performed at low-cost and by repurposing existing equipment~\cite{whitepaper}. They have the potential to improve upon the MTA Atomki experiment by incorporating high-resolution HPGe detectors, allowing for $\lesssim 70$ keV resolution in reconstructing the electron-positron invariant mass (see Ref.~\cite{whitepaper} for a more detailed discussion of these proposals).

This brief study emphasizes a broader science case for new physics searches in rare nuclear transitions utilizing high-resolution detectors.  Our main observations are that given sufficient luminosity, nuclear transition experiments have the potential to compete with other upcoming fixed target and collider experiments in their sensitivity to new physics in the $\sim 1-20$ MeV range, and provide a unique probe of the hadronic and leptonic couplings of light hidden particles, whereas the vast majority of upcoming experiments will only be sensitive to leptonic couplings in this mass range. Of course, these searches would also definitively scrutinize the existence of a new $\sim17$ MeV particle as suggested by the MTA Atomki experimental results. On a practical level, an additional attractive feature of these experiments is that they are cost-effective and could begin collecting data on a short timescale.

To illustrate the potential for rare nuclear transition searches to competitively probe new physics, we study the sensitivity of follow-up $^8$Be experiments to the well-known dark photon scenario, in which a new light vector gauge boson, $A^{\prime}$, couples to the Standard Model electromagnetic current through kinetic mixing, $\varepsilon$, with hypercharge.  Although it is only one of many possible models for light new MeV-scale physics, it provides a useful benchmark for comparing experimental sensitivities. We will show that experiments incorporating high-resolution HPGe detectors and integrated luminosity similar to that of the MTA Atomki experiment can probe currently unexplored regions of the dark photon parameter space.

\section{Dark Photons in $^8$Be Transitions}\label{sec:crosssections}

We consider the production and subsequent decays of excited states of $\beg$ (see Ref.~\cite{Tilley:2004zz} for the detailed properties of this system). The states of interest, denoted $\bet$ and $\bes$, lie at 17.64, 18.15 MeV above the ground state (denoted by $\beg$). Both have $J^P$ quantum numbers $1^+$, while the ground state is $0^+$.  These resonances are admixtures of isospin eigenstates, $\bes$ being predominantly isoscalar and $\bet$ mostly isovector. The lower-lying $\bet$ state is significantly narrower, with $\Gamma \simeq 10.7 $ keV, while the $\bes$ width is $\Gamma \simeq 138 $ keV. When produced, $\bet$ and $\bes$ primarily decay back to $^7$Li$+p$, but can also decay radiatively through a photon to the $^8$Be ground state, with~\cite{Tilley:2004zz} 
\beq \label{eq:BR}
BR(\bet \rightarrow \beg \, + \gamma) \approx 1.4 \times 10^{-3}, \quad  BR(\bes \rightarrow \beg \, + \gamma) \approx 1.4 \times 10^{-5}. 
\eeq
Note the substantially larger branching ratio for electromagnetic $\bet$ decay. More rarely, $\bet$ and $\bes$ can de-excite to the ground state via ``internal pair creation'' (IPC)~\cite{Rose:1949zz}, whereby an electron-positron pair is produced via an off-shell photon. The $e^+ e^-$ branching ratios are predicted to be $\sim 3 \times 10^{-3} BR(\gamma)$~\cite{Rose:1949zz} for both $\bet$ and $\bes$, where $BR(\gamma)$ is the corresponding branching ratio in Eq.~(\ref{eq:BR}). Decays of these states to $\beg$ via real photon emission are $M1$ transitions. Note that while we focus exclusively on $\bes$ and $\bet$ in this work, there are several other excited states and decay channels within $\sim 20$ MeV of the $\beg$ ground state~\cite{Tilley:2004zz}.

Both $\bes$ and $\bet$ can be resonantly produced by bombarding a $^7$Li target with a proton beam. The proton energies required for populating these excited states are $E_p = 440$ keV, $1.03$ MeV for  $\bet$ and $\bes$, respectively. This is the method used by the MTA Atomki experiments, which target the processes
\beq
\begin{aligned}
&p + \lig \rightarrow \bes \rightarrow e^+ \, e^{-}  \, +\beg\\
&p + \lig\rightarrow \bet \rightarrow e^+ \, e^{-}  \, + \beg.
\end{aligned}
\eeq
These transitions can receive contributions from new hidden particles. For example, a dark photon $A^{\prime}$ with mass $m_{A^{\prime}}\lesssim 17$ MeV could be produced on-shell from the decay of either $\bet$ or $\bes$, and subsequently decay to $e^+ e^-$. The Atomki experimental results in Ref.~\cite{Krasznahorkay:2015iga} were interpreted as evidence for a contribution of this type in $\bes$ and, more recently~\cite{Krasznahorkay:2017qfd, Krasznahorkay:2017gwn}, $\bet$ transitions (although this cannot be explained by a dark photon with kinetic mixing alone due to existing constraints from other experiments~\cite{Feng:2016jff, Feng:2016ysn}). Internal pair creation via an off-shell photon becomes an irreducible background for these searches. 

Our goal is to investigate the extent to which future experiments can probe the canonical dark photon parameter space. To model both the dark photon and standard IPC contributions to the above processes, we make use of the effective field theory description for the $\beg$ system recently formulated in Ref.~\cite{Zhang:2017zap}. This approach allows one to include angular dependence and interference effects in predictions for $p+\lig \rightarrow \beg \,+ e^+ \, e^-$ production. In this method, the relevant couplings are determined from data for the strong decay widths and rates for the pure electromagnetic transitions to on-shell photons in Eq.~(\ref{eq:BR}). 

The production cross-section for $p + \lig \rightarrow  \beg \, + \gamma$ (averaged over initial spin states) can be computed from~\cite{Hammer:2017tjm}
\beq \label{eq:dsdOphoton}
\frac{d \sigma}{d \Omega} = \frac{\mu \,  q}{64\,\pi^2 \,p} \sum_{a, \sigma, \lambda} \left| \epsilon_{\mu}^* \mathcal{M}^{\mu} \right|^2
\eeq
where $p$ and $q$ are the momenta of the incoming proton and outgoing photon, $\mu$ is the reduced mass of the $p - ^7{\rm Li}$ system, $a$ and $\sigma$ are the $^7$Li and proton spin projections, $\lambda$ is the helicity of the outgoing photon with photon polarization vector $\epsilon_{\mu}^*$, and  $\mathcal{M}^\mu$ is the matrix element
\beq
 \mathcal{M}^\mu \equiv  \langle  \beg \, + \gamma | J_{\rm EM}^{\mu} | ^7{\rm Li} + p  \rangle ,
\eeq
with $J_{\rm EM}^{\mu}$ the electromagnetic current. The components of $\mathcal{M}^\mu$ are computed in Ref.~\cite{Zhang:2017zap} in the halo effective field theory approach and incorporating Coulomb effects in the initial scattering state. All kinematic quantities above are in the $p - ^7{\rm Li}$ center-of-mass (COM) frame. The photon polarization vectors are most conveniently expressed in a helicity basis, whereby the spatial components are labeled by quantum numbers $j=-1,0,+1$ for the projection of the total angular momentum along the quantization axis. In this basis, the polarization vectors for a massless photon are
\beq
(\epsilon^{*}_{\lambda})^{t} =  0, \quad (\epsilon_{\lambda}^{*})^0 = 0, \quad (\epsilon_{\lambda}^{*})^{\pm 1} = \delta_{\lambda \pm 1}
\eeq
where $\lambda=-1, 0,+1$. In the conventions and notation of Ref.~\cite{Zhang:2017zap}, the product  $ \epsilon_{\mu}^* \mathcal{M}^{\mu} $ for a given set of spins and helicity $\lambda$ is
\beq \label{eq:components}
 (\epsilon_{\lambda}^*)_{\mu} \mathcal{M}^{\mu} = (\epsilon_{\lambda}^*)^{\mu} \mathcal{M}_{\mu} = (\epsilon_{\lambda}^*)^t J_t - (\epsilon_{\lambda}^*)^j J_j
\eeq
with $j = -1,0,+1$ and the components $J_t (= J^t)$, $J_j$ as in Eq.~(3.1) of Ref.~\cite{Zhang:2017zap}. 

\begin{figure}[t!]
\begin{center}
\includegraphics[width=0.65\linewidth]{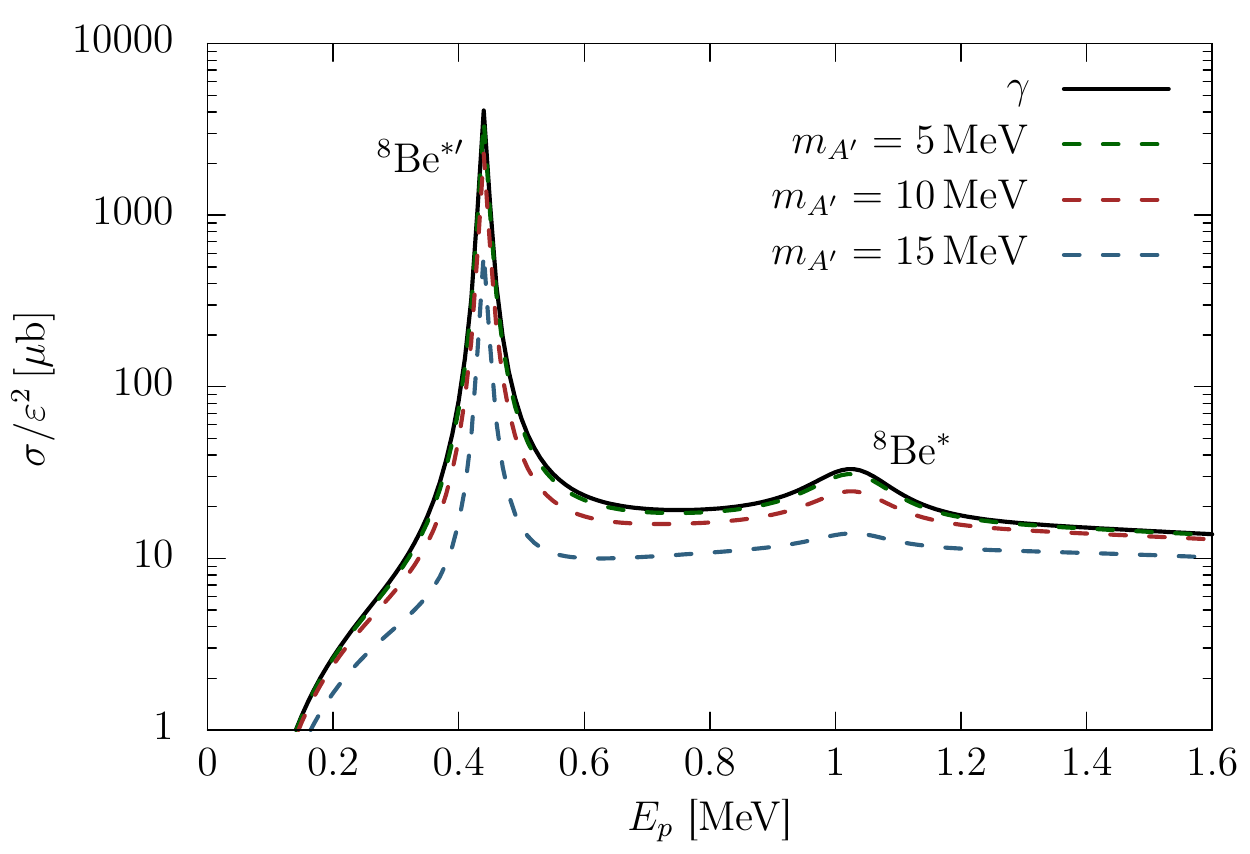}
\caption{Total cross-sections for $p + \lig \rightarrow  \beg \, + \gamma$ (black) and $p + \lig \rightarrow  \beg \, + A^{\prime}$ for $m_{A^{\prime}}=5$, 10, 15 MeV (dashed curves) as a function of the beam energy and the kinetic mixing $\varepsilon^2$ factored out. The $\bet$ and $\bes$ resonances are clearly visible.} 
\label{fig:cs}
\end{center}
\end{figure} 

Utilizing the expressions above, we show results for the total total $p + \lig \rightarrow  \beg \, + \gamma$ cross-section for different beam energies $E_p$ in Fig.~\ref{fig:cs}. This treatment accounts for both the resonant $M1$ contributions involving $\bet$ and $\bes$, as well as the $E1$ and $E2$ contributions from non-resonant proton capture. The $\bet$ and $\bes$ resonances are clearly visible and the results agree well with experimental data~\cite{Zhang:2017zap, Zahnow1995}.

The expressions above can be straightforwardly modified to account for the emission of an on-shell massive dark photon. In this case, Eq.~(\ref{eq:dsdOphoton}) can again be used, as can Eq.~(\ref{eq:components}) and the expressions for the components of the matrix element in Eq.~(3.1) of Ref.~\cite{Zhang:2017zap}, with an overall rescaling by $\varepsilon$ ($\varepsilon^2$ in the cross-section). Now however, one must  account for the longitudinal polarization of the dark photon\footnote{Note that, for a pure $M1$ transition in the center of mass frame, the contribution from longitudinal emission vanishes by conservation of angular momentum and parity.}. The corresponding polarization vectors are
\beq
(\epsilon^{*}_{\lambda})^{t} =  \frac{q}{m_{A}} \delta_{\lambda 0}, \quad (\epsilon_{\lambda}^{*})^0 = \frac{\omega}{m_{A^{\prime}}} \delta_{\lambda 0}, \quad (\epsilon_{\lambda}^{*})^{\pm 1} = \delta_{\lambda \pm 1}
\eeq
where $q$ is the dark photon momentum and $\omega$ its energy. Results for the total $p + \lig\rightarrow  \beg \, + A^{\prime}$ cross-section are also shown in Fig.~\ref{fig:cs}, with $\varepsilon^2$ factored out and for different masses. The cross-sections for the massive vector are suppressed relative to the $\gamma$ transition strength by $\varepsilon^2$ and by powers of $|\mathbf{p}_{A^{\prime}}/|\mathbf{p}_{\gamma}|$, as can be seen from the expressions of Ref.~\cite{Zhang:2017zap} (this suppression for the $M1$ contribution is also discussed in Refs.~\cite{Feng:2016jff, Feng:2016ysn}).

We also require the cross-section for the IPC background, $p + \lig \rightarrow  \beg \, + \gamma^* \rightarrow \beg \, + e^+ \, e^-$. The differential cross-section for this process in the Standard Model can be written as~\cite{Zhang:2017zap}
\beq \label{eq:dsdOepem}
\frac{d \sigma}{d  E_+ d\cos \theta_{\pm} d \cos\theta d\phi} = \frac{\mu \,\alpha_{\rm EM} \, p_+ \, p_-}{128 \pi^3 p} \sum_{\rm spins} \left| \mathcal{M} \right|^2,
\eeq
where the $\pm$ subscripts refer to the positron/electron, $\theta_{\pm}$ is the $e^+ e^-$ opening angle, and $\phi$ is the angle between the electron-postron plane and the plane defined by the beam axis and the (virtual) photon momentum (see Ref.~\cite{Zhang:2017zap} for a more detailed discussion of this setup). All quantities are again understood to be defined in the  $p - \lig$ COM frame. The matrix element appearing in Eq.~(\ref{eq:dsdOepem}) is specified by Eqs.~(4.1)-(4.8) in Ref.~\cite{Zhang:2017zap}. The differential cross-section above can be re-expressed in terms of the invariant mass of the electron-positron pair,
\beq \label{eq:Mpm}
M_{\pm}^2 = E_+^2 + \left(\omega -E_+\right)^2 - 2\sqrt{\left(E_+^2-m_e^2\right)\left(\left(\omega - E_+\right)^2 - m_e^2\right)}\cos \theta_{\pm}.
\eeq
The differential cross-section $d \sigma /d M_{\pm}$, marginalized over the angular ranges appropriate for a given detector configuration, can be used to search for peaks from new particle decays to $e^+ e^-$ pairs, as discussed below. The IPC prediction from Eq.~(\ref{eq:dsdOepem}) provides a good match to the Atomki data away from the reported anomaly, once detector efficiencies are accounted for~\cite{Krasznahorkay:2015iga, Zhang:2017zap, Gulyas:2015mia}.

\section{Experimental Sensitivities}

With the expressions above, we can estimate the experimental sensitivities to dark photon production via decays of $\bet$ and $\bes$ and compare to the projected sensitivities of other experiments.

Let us assume an experimental setup similar to that utilized by the MTA Atomki group~\cite{Gulyas:2015mia}. We consider a circular detector set up over the target in a plane perpendicular to the beam-line. For simplicity, we will assume that only events with $\theta \simeq \pi/2$ and $\phi\simeq \pi/2, \, 3\pi/2$ will be observed by the detector. The experiment would detect the $e^+ e^-$ individual energies\footnote{For the high mass resolutions we consider below, this is nontrivial to achieve and will likely require particle identification.} and angle between them, $\theta_{\pm}$, and use this information to reconstruct the invariant mass of the electron-positron pair using Eq.~(\ref{eq:Mpm}) above. The $M_{\pm}$ distribution can then be scanned for a bump-like feature, corresponding to the peak expected from the decay of a massive particle.

\begin{figure}[!t]
\begin{center}
\includegraphics[width=0.55\linewidth]{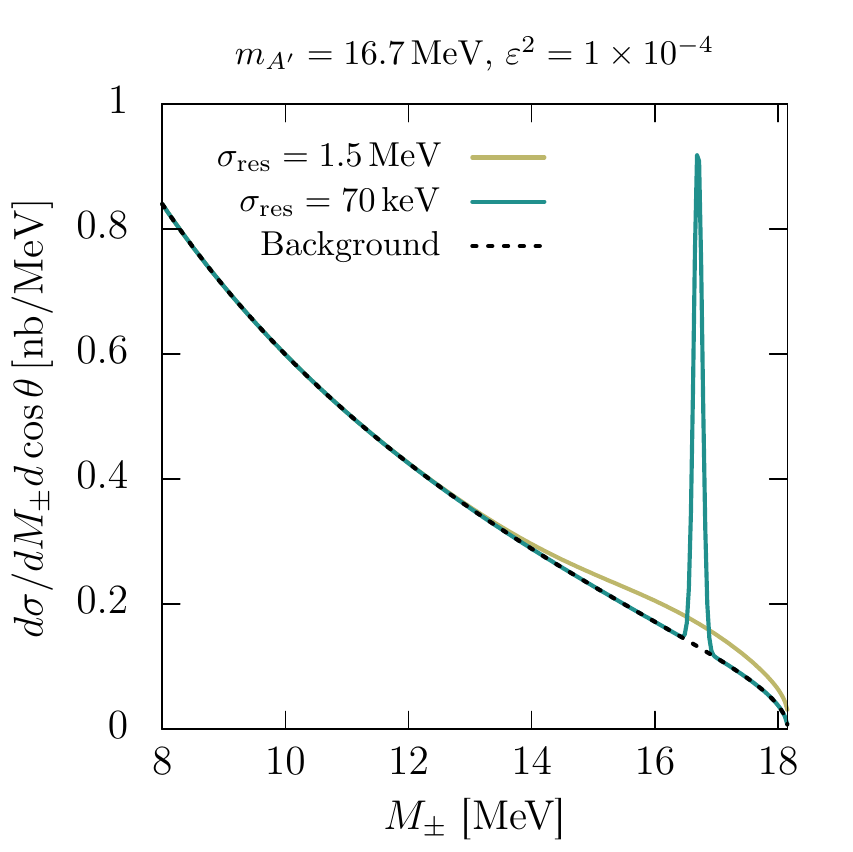}
\caption{Invariant mass distributions for $p + \lig \rightarrow \beg \, + e^+ \, e^-$, including contributions from both the standard IPC background (black dashed curve), and from a dark photon with $m_{A^{\prime}} = 16.7$ MeV, $\varepsilon^2 = 1\times10^{-4}$ with beam energy $E_p=1.03$ MeV. This parameter space point is roughly consistent with the Atomki signal~\cite{Feng:2016jff}, although it is excluded by existing measurements in the pure kinetic mixing scenario. The yellow curve shows the prediction for mass resolution $\sigma_{\rm res}=1.5$ MeV. The green curve corresponds to the prediction for $\sigma_{\rm res}=70$ keV. Improvements in mass resolution will have an important impact on the sensitivities of future experiments.} 
\label{fig:dist}
\end{center}
\end{figure} 

We will make use of the narrow width approximation (NWA), whereby, for a given beam energy, we estimate the corresponding invariant mass distribution as the sum of the standard IPC result derived from Eq.~(\ref{eq:dsdOepem}) and a contribution from the decay of an on-shell dark photon. The latter is taken to be a Gaussian centered around $M_{\pm} = m_{A^{\prime}}$, normalized to the prediction of Eq.~(\ref{eq:dsdOphoton}) at the relevant angles, and with width $\sigma_{\rm res}$ set by the experimental mass resolution.  We assume $BR(A^{\prime}\rightarrow e^+ e^-) \approx 1$. The decay width of the dark photon in this mass range assuming only visible decays, $\Gamma \sim \alpha_{\rm EM} m_{A^{\prime}} \varepsilon^2$, is much smaller than the detector mass resolutions we consider, and so can be safely neglected. The NWA should reproduce the full distributions reasonably well, since $\Gamma/m_{A^{\prime}}\sim \alpha_{\rm EM} \varepsilon^2 \ll 1$ and the interference between the Standard Model and dark photon terms is suppressed by $\sim \alpha_{\rm EM}	$ relative to the the pure dark photon contribution to the cross-sections near the mass shell.  Note that, despite the small width, $A^{\prime}$ decays within $\lesssim 10$ mm of the production point across the parameter space we consider, and so would be visible in the $\sim$ cm - m--scale nuclear transition experiments of interest. 

The predicted invariant mass distribution for $m_{A^{\prime}}=16.7$ MeV and $\varepsilon^2 = 1\times 10^{-4}$ is shown in Fig.~\ref{fig:dist}. These particular parameters are excluded by other experiments, but are roughly those required in the dark photon scenario to explain the Atomki anomaly~\cite{Feng:2016jff}. Results are shown  for two different experimental mass resolutions: the yellow curve shows the prediction for $\sigma_{\rm res}=1.5$ MeV. This is similar to the resolution achieved by the Atomki experiments~\cite{Krasznahorkay:2015iga, Gulyas:2015mia}. The green curve corresponds to the prediction for $\sigma_{\rm res}=70$ keV, which is expected to be achievable by future experiments incorporating HPGe detectors~\cite{whitepaper}. Clearly, the mass resolution incorporated by future experiments will be a key factor in their sensitivity to new physics.

To estimate the extent to which future $\beg$ experiments could probe light new particles, we employ the following strategy: for a given $m_{A^{\prime}}$, we consider a window in $M_{\pm}$ centered on $m_{A^{\prime}}$ and with width $2\sigma_{\rm res}$ on each side of the central value. We then compute the expected number of signal ($S$) and background ($B$) events for a given luminosity in this window, utilizing the machinery presented above. We consider a particular mass and kinetic mixing to be observable if $S/\sqrt{B} \geq 3$. Of course this is a rather simplified analysis, but given the large background it should provide a reasonable preliminary estimate of the reach, neglecting the effects of systematic uncertainties. Note that, in the Atomki analysis, an additional cut on the parameter $y\equiv (E_- - E_+)/(E_- + E_+)$ was used to increase $S/B$. We do not impose this requirement in our treatment, but it may result in even better discrimination between signal and background.

In performing several searches across a large invariant mass range, significant statistical fluctuations in the background that mimic a signal are bound to occur. In a realistic setting one must account for this so-called ``look elsewhere effect'' to assign global significance to an observation. For example, if the total invariant mass range considered is $[2\,m_e, \,17.5$ MeV$]$ and a resonance search is done in $~2 \sigma_{\rm res}=140$ keV non-overlapping (double-sided) windows, this corresponds to testing $\sim 60$ hypotheses (more masses will likely be considered in a realistic experiment). If we require $\sim 95$\% C.L.~globally to claim sensitivity, the look-elsewhere correction would roughly correspond to requiring local $p$-values smaller than $\sim 0.05/60$. This motivated our choice of requiring $S/\sqrt{B}\geq 3$ rather than $\geq 2$, which is more frequently used.

\begin{figure}[t!]
\begin{center}
\includegraphics[width=0.65\linewidth]{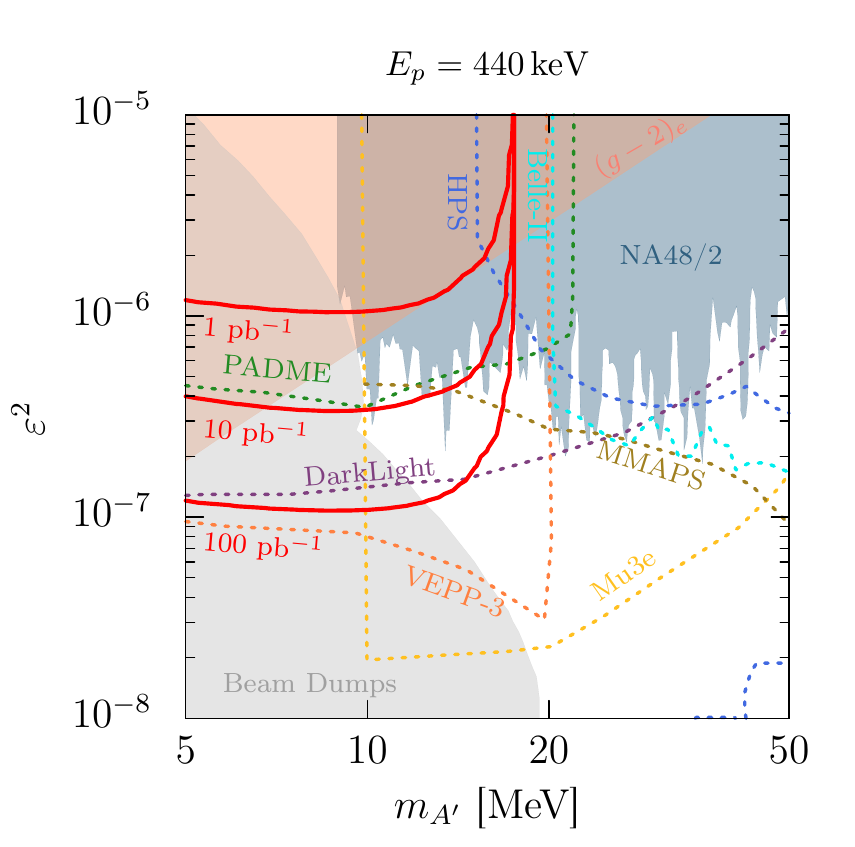}
\caption{Projected sensitivities for different effective integrated luminosities, $L_{\rm eff}$ across the dark photon parameter space for a $^8$Be nuclear transition experiment with $E_p=440$ keV and mass resolution $\sigma_{\rm res}=70$ keV. Also shown are current exclusion limits (shaded) and projections (dashed) of other experiments that are expected to have results by 2021 (adapted from Refs.~\cite{whitepaper, Alexander:2016aln}). High-resolution nuclear transition experiments can begin to cover unexplored regions of the parameter space with $L_{\rm eff} \gtrsim 2$ pb$^{-1}$. For reference, the MTA Atomki $^8$Be$^{*}$  experiment achieved $\sim \mathcal{O}(1$ pb$^{-1})$~\cite{Krasznahorkay:2015iga}. } 
\label{fig:projections}
\end{center}
\end{figure} 

The approximate expected sensitivity, as defined above, for an experiment with beam energy $E_p=440$ keV (populating the $\bet$ resonance) and mass resolution $\sigma_{\rm res}=70$ keV is shown in Fig.~\ref{fig:projections} for several different effective integrated luminosities, $L_{\rm eff}$, defined by
\beq
L_{\rm eff} \equiv \tilde{\epsilon} \times \int \mathcal{L} dt.
\eeq
Here $\mathcal{L}$ the instantaneous luminosity and $\tilde{\epsilon}$ an approximate acceptance factor times detector efficiency (assumed to be flat across energies and angles). For reference, $L_{\rm eff}$ for the Atomki experiment sitting on the $\bes$ resonance presented in Ref.~\cite{Krasznahorkay:2015iga} appears to be $\mathcal{O}(1$ pb$^{-1})$. For a fixed $L_{\rm eff}$, the sensitivity to dark photons from observations of $\bet$ decays is expected to be considerably better than that from $\bes$ for $m_{A^{\prime}}\lesssim 17$ MeV. This can be seen from Eq.~(\ref{eq:BR}): for fixed $m_{A^{\prime}}$, $\varepsilon$, the dark photon branching ratio is proportional to the $\beg \, + \gamma$ partial width, which is two orders of magnitude larger for $\bet$. This is also evident from Fig.~\ref{fig:cs}.

The results shown in Fig.~\ref{fig:projections} agree with a simple back-of-the envelope estimate for the projected sensitivities. The Atomki experiment is roughly sensitive to $\varepsilon^2\sim 10^{-4}$~\cite{Feng:2016jff, Feng:2016ysn}. Improving the mass resolution from $\sim 1.5$ MeV to $\sim 70$ keV results in a reduction of $\sqrt{B}$ by a factor of $\sim 1/5$. Meanwhile, both the signal and background cross-sections are roughly $\sim 100$ times larger sitting on the $\bet$ resonance than on the $\bes$ resonance (c.f.~Fig.~\ref{fig:cs}), which increases $S/\sqrt{B}$ by a factor of $\sim 10$ for fixed $m_{A^{\prime}}$, $\varepsilon^2$. Taken together, this suggests that such an experiment could have sensitivity to $\varepsilon^2$ values roughy a factor of $50$ smaller than the Atomki setup given the same $L_{\rm eff}$, which corresponds to $\varepsilon^2\sim 10^{-6}$. This agrees with the 1 pb$^{-1}$ curve in Fig.~\ref{fig:projections}.

The sensitivity projections above neglect the effects of systematic uncertainties and assume that sources of background besides the standard IPC contribution can be effectively eliminated. This may be challenging to achieve in a realistic counting experiment. While we are not able to estimate the level of systematic uncertainties that can be attained by such experiments, we note that $S/B \gtrsim 0.1-1\%$ in $2\sigma_{\rm res}=140$ keV windows across the $\varepsilon^2 \gtrsim 10^{-7}$ region. Sensitivity with this level of signal purity does not appear to be unreasonable compared to what can be achieved by other experiments~\cite{Alexander:2016aln}, and it could be possible to design an analysis that mitigates the effects of these uncertainties on the sensitivity to new physics. For example, uncertainties in the predicted number of background events could be reduced by considering sidebands in the quantity $y=(E_- - E_+)/(E_- + E_+)$ which are not expected to be populated by the signal (see e.g.~\cite{Krasznahorkay:2015iga}). 

To evaluate the impact of a potentially mis-modeled background on our sensitivity projections, we also perform a simple background-agnostic analysis typical of a ``bump hunt''. For a given $m_{A^{\prime}}$, we generate a sample of Monte Carlo pseudo-experiments, simulating the background + signal distributions assuming Poisson statistics for each bin. We then scan over the resulting distributions in non-overlapping $2\sigma_{\rm res}$ windows. In each window, we fit a line and a line + Gaussian to the simulated distribution and construct the maximum likelihood ratio. This yields a local median $p$-value for discovery (rejecting the no-signal hypothesis). Requiring the median local $p$-value to be less than about $10^{-3}$ (which should correspond roughly to a $2\sigma$ global deviation from the expected background), yields sensitivity estimates for each integrated luminosity that match up quite well with those of the $S/\sqrt{B}$ analysis. This suggests that the projections shown in Fig.~\ref{fig:projections}, obtained from a simple cut-and-count analysis, are also indicative of the sensitivity that can be achieved by a data-driven bump hunt, provided the full experimental background varies slowly across $2\sigma_{\rm res}$ windows, and that its magnitude is not significantly larger than that predicted by the expressions in Sec.~\ref{sec:crosssections}. We expect the latter method to be significantly less susceptible to the effects of systematic uncertainties.

In Fig.~\ref{fig:projections} we also show current exclusion limits from various experiments. The most stringent in the 1-20 MeV mass range are from beam dump experiments (gray), $(g-2)_e$ measurements (orange) and the NA48/2 experiment (blue) at CERN. Note that of these, only the NA48/2 experiment is sensitive to hadronic couplings. The bounds shown are adapted from Refs.~\cite{whitepaper, Alexander:2016aln}, but including the updated NA48/2 limits of Ref.~\cite{Lurkin:2017aqo}. 

Comparing our projected sensitivities with current limits suggests that an experiment targeting the $\bet$ resonance with 70 keV mass resolution and integrated luminosity roughly similar to that achieved in the MTA Atomki $^8$Be$^{*}$ experiment~\cite{Krasznahorkay:2015iga} could begin to cover currently unexplored regions of the parameter space. It should also be noted that the exclusions from $(g-2)_e$ measurements, which dominate for $m_{A^{\prime}} \sim 1- 10$ MeV and $\varepsilon^2$ above the beam dump limits, are indirect and model-dependent, in that they do not directly search for new hidden particles themselves. Nuclear transition experiments, even with relatively low integrated luminosity, could provide one of the first \emph{direct} probes of many scenarios with hidden particles in the $\sim 1 - 10$ MeV mass range. In fact, the existing Atomki results (away from the anomaly) could likely be used to extract dark photon limits down to $ \varepsilon^2 \sim 10^{-4}$ or so, although their detector efficiencies decrease rapidly for the small angles relevant for $m_{A^{\prime}} \lesssim 10$ MeV~\cite{Gulyas:2015mia}.

There are many experiments, either currently underway or in the design stages, that are expected to probe the parameter space accessible to nuclear transition experiments before 2021~\cite{Balewski:2013oza, Balewski:2014pxa, Wojtsekhowski:2009vz, Wojtsekhowski:2012zq, Raggi:2014zpa, Blondel:2013ia, Echenard:2014lma, Denig:2016dqo, Battaglieri:2014hga, TheBelle:2015mwa}. A sample of these are shown in Fig.~\ref{fig:projections}. The projections are again adapted from Refs.~\cite{whitepaper, Alexander:2016aln}. Note that  projections for the currently running NA64 experiment at CERN are not shown in Fig.~\ref{fig:projections}. Depending on beam time and energy, NA64 will likely also probe a significant portion of the parameter space shown in the 1-2 year time scale~\cite{Gninenko:2013rka,Andreas:2013lya}. 

The results in Fig.~\ref{fig:projections} show that, provided sufficient luminosity, nuclear transition experiments investigating the Atomki anomaly in $\beg$ can be competitive with other fixed target and collider probes of the dark photon parameter space in the 1-20 MeV mass range. Such an experiment, once funded, could realistically be designed, constructed, and begin to take data within 1-2 years, possibly reaching the targeted luminosities shown in Fig.~\ref{fig:projections} by 2020 or 2021~\cite{lang}. 

In the meantime, several of the experiments shown on Fig.~\ref{fig:projections} will likely also produce results, and so one may question the usefulness of an additional probe of what appears to be the same parameter space. It is important to realize, however, that the nuclear transition experiments discussed here probe couplings of new hidden particles to both leptons \emph{and} hadrons. In the dark photon scenario, these couplings are related by the relative electric charges, but this is not the case in general new physics scenarios (see e.g.~\cite{Feng:2016ysn, Kahn:2016vjr, Kozaczuk:2016nma} for some examples). To our knowledge, all of the other experiments projected to probe the parameter space shown within the next few years will be exclusively sensitive to leptonic couplings, and so rare nuclear transition experiments would be complementary to these existing proposals. 

While we have focused on experiments targeting $^8$Be, other nuclear systems could provide comparable or better reach. For example, $^4$He would likely feature a cleaner environment and provide sensitivity up to masses between 20-30 MeV.  The lowest-lying $0^+$ excited state of this system is at $\sim 20$ MeV above the $0^+$ ground state, with several other $J=1$ resonances in the 20-30 MeV range~\cite{Tilley:1992zz}. It would be interesting to perform a more detailed study of the prospects for observing light bosons in $^4{\rm He}$ transitions in the future.

\section{Conclusions}

We have argued that experiments analyzing rare nuclear transitions with high-resolution detectors can be competitive with other fixed-target and collider experiments in probing MeV-scale new physics. In addition, they would definitively scrutinize the $^8$Be anomaly, provide sensitivity to both hadronic and leptonic couplings of light hidden particles, and yield useful information about the nuclear properties of the corresponding systems. Taken together, these observations provide a compelling case for seriously pursuing such experiments in the near future.

\section*{Acknowledgements}

I'm grateful to Jonathan Feng, Iftah Galon, and Xilin Zhang for stimulating discussions on this topic, as well as the other participants of the U.S. Cosmic Visions: New Ideas in Dark Matter workshop, where this project was initiated. I also thank Attila Krasznahorkay for clarifications about the Atomki experimental analysis, as well as David E.~Morrissey for useful feedback on this manuscript. I am particularly grateful to Patrick Draper, David Koltick, and Rafael Lang for many invaluable conversations and collaboration at various stages of this project, as well as feedback on this manuscript. 

\bibliographystyle{JHEP}
\bibliography{nuclear_transitions}

\end{document}